\documentclass[11pt]{article}
\usepackage[utf8]{inputenc}
\usepackage{amsmath}
\usepackage{amsfonts}
\usepackage{amssymb}
\usepackage{jcappub}
\usepackage{graphicx}
\usepackage{placeins}
\usepackage{float}
\usepackage{breqn}
\usepackage{soul}

\begin{document}

\author[a]{Nina K. Stein}
\emailAdd{ninastei@buffalo.edu}
\author[a]{William H. Kinney}
\emailAdd{whkinney@buffalo.edu}
\affiliation[a]{Univ. at Buffalo, SUNY, Buffalo NY USA}
\title{Natural Inflation After Planck 2018}
\linespread{2.5}

\abstract{
We calculate high-precision constraints on Natural Inflation relative to current observational constraints from Planck 2018 + BICEP/Keck(BK15) Polarization + BAO on $r$ and $n_S$, including post-inflationary history of the universe. We find that, for conventional post-inflationary dynamics, Natural Inflation with a cosine potential is disfavored at greater than 95\% confidence out by current data. If we assume protracted reheating characterized by  $\overline{w}>1/3,$ Natural Inflation can be brought into agreement with current observational constraints. However, bringing unmodified Natural Inflation into the 68\% confidence region requires values of $T_{\mathrm{re}}$ below the scale of electroweak symmetry breaking. The addition of a SHOES prior on the Hubble Constant $H_0$ only worsens the fit. 
}
\maketitle

\section{Introduction}
Inflationary cosmology, first proposed by Guth in 1980 \cite{Starobinsky:1980te,Sato:1981ds,Sato:1980yn,Kazanas:1980tx,Guth:1980zm,Linde:1981mu,Albrecht:1982wi}, remains an widely studied and successful approach for understanding the early universe, solving at a stroke the flatness and horizon problems, as well as providing a mechanism to generate the observed primordial power spectrum \cite{Starobinsky:1979ty,Mukhanov:1981xt,Mukhanov:2003xw,Linde:1983gd,Hawking:1982cz,Hawking:1982my,Starobinsky:1982ee,Guth:1982ec,Bardeen:1983qw}. Inflation relates the evolution of the universe to one or more scalar \textit{inflaton} fields, the properties of which dictate the dynamics of the period of rapidly accelerating expansion which terminates locally in a period of reheating, followed by radiation-dominated expansion. While we cannot precisely determine specific form of the potential for the inflaton field or field, different choices of potential result in different values for cosmological parameters, which are distinguishable by observation \cite{Dodelson:1997hr,Kinney:1998md}. Recent data, in particular the Planck measurement of Cosmic Microwave Background (CMB) anisotropy and polarization \cite{Ade:2015lrj,Ade:2015xua,Aghanim:2015xee,Aghanim:2018eyx,Aghanim:2019ame,Akrami:2019bkn,Akrami:2019izv,Akrami:2018odb}  and the BICEP/Keck measurement of CMB polarization \cite{Ade:2015fwj,Ade:2015tva} now place strong constraints on the inflationary parameter space, falsifying many previously viable inflationary potentials, including some of the simplest and most theoretically attractive models. 

One such class of models is Natural Inflation, put forward in 1990 by Freese, Frieman and Olinto \cite{Freese:1990rb} as a solution to certain theoretical challenges inherent to slow rolling inflation models, which are limited by the fact that in order to generate an adequate amount of inflation, the slope of the inflaton potential must be very nearly flat. This creates fine-tuning problems, in particular, quantum corrections in the absence of a symmetry generically spoil the flatness of the potential, which is known as the \textit{$\eta$-problem}.
Natural inflation (NI) models avoid this by using an axionic field to drive inflation, where the term ``axionic'' refers in the most general sense to a field which has a flat potential as a result of a shift symmetry. During the early Universe, explicit breaking of the shift symmetry gives rise to slow-roll expansion In this sense the inflaton in NI is a  pseudo-Nambu-Goldstone  boson, with a nearly flat potential, exactly as inflation requires.

When Freese, Frieman and Olinto proposed their original model of Natural Inflation in 1990, they modeled the inflaton field directly on the QCD axion, albeit with a different mass scale. As with the QCD axion, the potential was an ordinary cosine, with a height of $\approx 10^{16}\ \mathrm{GeV}$ and a width of at least $10^{19}\  \mathrm{GeV}$, to match CMB observations \cite{Freese:2014nla,Freese:1990rb}. Subsequently, many other variants have been proposed, such as axion monodromy \cite{Silverstein:2008sg,Kobayashi:2014ooa,Higaki:2014sja}, but for the purposes of this paper, we shall confine ourselves to discussing the original cosine potential, though the principle can be extended to cover many other potentials. 

The two primary observable parameters of the primordial power spectrum (as seen in the CMB) used to determine the viability of inflationary models are: 1. $r$, the ratio of tensor (gravitational wave) to scalar (density) perturbations,  ($r\equiv P_T/P_R$), and 2. $n_s$, the spectral index, which describes the degree of scale dependence of the fluctuation amplitude. While neither of these carry any direct dependence on post-inflationary dynamics, they do depend on the value of $N_k$, the number of e-folds of expansion between the point when fluctuation modes on the pivot scale (generally taken to be $k=0.002\ \mathrm{Mpc}^{-1}$) exited the horizon and the end of inflation. This is typically around 60 e-folds, but the actual value depends on the evolution of the universe between the end of inflation and nucleosynthesis, a dependence we explain in some detail in Sec. 2.2 of this paper. We find that, assuming conventional post-inflationary dynamics, current observations entirely rule out standard Natural Inflation, but that by positing a period between the end of inflation and nucleosynthesis during which the universe expands at a rate slower than radiation domination, we can bring it back into agreement with the data. (Other modifications to improve agreement with data have been proposed, for example thermal dissipative effects \cite{Reyimuaji:2020bkm} and non-minimal coupling to gravity \cite{Reyimuaji:2020goi}.)

The structure of this paper is as follows: in Sec. 2, we cover inflationary theory, first in terms of the mechanics of Natural Inflation, and then explain how the inflationary epoch parameters relate to modern-day observables; in Sec. 3, we discuss the way the reheating parameters influence these observables, and then the methodology of our calculations; Sec. 4 presents our results, first in the conventional case and then in the more general scenario, and we present our conclusions in Sec. 5.

\section{Theory}

\subsection{Natural Inflation}
In order to generate sufficient inflation while still satisfying observational constraints on anisotropy in the cosmological microwave background (CMB), the inflaton field must be characterized by an extremely flat potential; assuming a single-field inflationary model, the ratio of the potential's height to its width must satisfy \cite{Adams:1990pn} 
\begin{equation}
\chi \equiv \frac{\Delta V}{\left(\Delta \phi \right)^4}\leq \mathcal{O}\left(10^{-6}-10^{-8}\right),
\label{nim1}
\end{equation}
where $\Delta V$ is the change in the inflationary potential $V(\phi)$ and $\Delta \phi$ is the change in the inflaton field $\phi$ during the slow roll portion of the inflationary period. The inflaton self-coupling must therefore be exceedingly weak, with an effective quartic self-coupling constant $\lambda_{\phi} <10^{-12}$ for reasonable models. \cite{Freese:2014nla} This extreme ratio between mass scales is referred to as the ``fine-tuning'' problem in inflation, a review of which can be found in Ref. \cite{Bassett:2005xm,Kinney:2009vz,Baumann:2009ds}. 

Natural Inflation approaches this problem by positing  that the inflaton potential is flat due to the presence of a shift symmetry, that is, $V(\phi)=V(\phi+\mathrm{const.})$  If the symmetry were perfect, such an inflaton could not roll and drive inflation, so we require a further explicit symmetry breaking, rendering the inflatons pseudo-Nambu Goldstone bosons (PNGBs) with the desired very nearly (but not \textit{exactly}) flat potential. Such a model naturally generates the small mass scale ratio specified in equation \ref{nim1}; for comparison, the QCD axion has a corresponding ratio of order $10^{-64}$, significantly smaller than inflation requires. \cite{Freese:2014nla} Furthermore, this ratio of scales is stable to radiative corrections because of the underlying global symmetry of the Lagrangian. 

The original NI model, which we address in this paper, is characterized by a potential of the form \cite{Freese:1990rb}
\begin{equation}
V(\phi)=\Lambda^4\left[1\pm \cos\left(N\phi/f\right)\right],
\end{equation}
where we will be considering the positive root, and defining $\frac{f}{N}\equiv\mu$, both of which can be done without loss of generality. Thus, the actual form of the potential we will be discussing is 
\begin{equation}
V(\phi)=\Lambda^4\left[1+ \cos\left(\phi/\mu\right)\right].
\label{nim2}
\end{equation}
Given appropriate scales for $\Lambda$ and $\mu$ ($\approx m_{\mathrm{GUT}}$ and $\approx m_{\mathrm{Pl}}$, respectively), such an inflaton potential can drive inflation, producing an appropriately small value of $\chi$ to satisfy \ref{nim1} with an inflaton mass of $m_{\phi}=\Lambda ^2 /\mu \approx \mathcal{O} \left( 10^{11}-10^{13}\ \mathrm{GeV}\right)$. \cite{Freese:2014nla}
\subsection{The Generation of Observables}
As the inflaton field rolls along the potential, quantum fluctuations generate perturbations in the metric, which rapidly increase in size until their wavelength exceeds the horizon size, at which point they `freeze out' and cease to evolve until they re-enter the horizon after inflation ends. The two primary types of perturbations are scalar modes, which represent fluctuations in density, and tensor modes, which represent gravitational wave fluctuations. The perturbation amplitude of these two types of fluctuations are given by 
\begin{equation}
    P_{\mathcal{R}}^{1/2}\left(k\right)=\frac{H^2\left(k\right)}{2\pi \dot{\phi}_k},
\end{equation}
for scalar modes, and 
\begin{equation}
     P_{\mathcal{T}}^{1/2}\left(k\right)=\frac{4H\left(k\right)}{\sqrt{\pi} m_{\mathrm{Pl}}},
\end{equation}
for tensor modes. In both cases, the left-hand side of the equation denotes the perturbation amplitude when a specific wavelength (pivot scale $k$) re-enters the Hubble radius after inflation, while the right-hand side is evaluated at the point during inflation where that same comoving wavelength froze out. \cite{Freese:2014nla} These two amplitudes are critical for evaluating the viability of inflationary models, as (under the slow roll approximation) $P_{\mathcal{R}}[k=0.002]\equiv A_s$ fixes the height of the inflationary potential, 
\begin{equation}
    A_s = \frac{H^2}{8 \pi^2 M_{\mathrm{P}}^2\epsilon}\bigg\vert_{k=aH},
\end{equation}
and the spectral index $n_s$ reflects the scale dependence of $P_{\mathcal{R}}$, 
\begin{equation}
    n_s-1 \equiv \frac{\mathrm{d}\ln P_{\mathcal{R}}}{\ln k},
\end{equation}
while the tensor amplitude is generally expressed in terms of the ratio between it and the scalar amplitude, $r\equiv P_{\mathcal{T}} / P_{\mathcal{R}}$. 

Since we use $A_s$ to normalize our potential, this leaves us $r$ and $n_s$ as observables whose values we can use to determine the viability of our models.

\FloatBarrier
\section{After Natural Inflation}

Standard inflationary cosmology assumes that at the end of inflation, the universe is characterized by a temperature of approximately $\mathcal{O}\left(10^{16}\ \mathrm{GeV}\right)$ and equation of state $w=-1/3,$ after which the universe undergoes a period of reheating, during which the inflaton field decays. Reheating can be instantaneous or protracted, and is generally defined by two parameters, the average equation of state, $\overline{w}$, and either the number of e-folds before the the universe enters a thermal equilibrium, radiation-domination epoch, $N_{\mathrm{re}}$, or, equivalently, the temperature at which this transition occurs, $T_{\mathrm{re}}$. This transition necessarily occurs before big bang nucleosynthesis (BBN), and therefore $T_{\mathrm{re}}>\mathcal{O}\left(1\ \mathrm{MeV}\right)$.\footnote{This is a rough estimate. For a more accurate treatment, see Refs. \cite{Kawasaki:2000en,Sabir:2019xwk}.} However, while BBN bounds place a lower bound on the onset of radiation domination at a temperature of $T_{\mathrm{re}}=1\ \mathrm{MeV}$,  unconventional dynamics below the scale of electroweak symmetry breaking at about $100\ \mathrm{GeV}$ could potentially have interesting implications for baryogenesis, and thus values of $T_{\mathrm{re}}$ below $100\ \mathrm{GeV}$ should be handled with some caution, particularly if the equation of state during reheating is greater than $1/3$ \cite{Cook:2015vqa,Tanin:2020qjw}.
Standard reheating assumes the weakly coupled decay of an oscillatory inflaton field at the end of inflation, and thus a reheating period characterized by $\overline{w}=0$, but many more complex models have been put forward \cite{Kofman:1997yn,Kofman:1994rk,Felder:2001kt,Felder:2000hj,Abolhasani:2009nb,Dufaux:2006ee,Shuhmaher:2005mf,Cook:2015vqa}. For such models, the average equation of state $\overline{w}$ can go as high as $\overline{w}=1,$ though any equation of state greater than 1 requires violating the dominant energy condition of general relativity/causality, and thus we only consider $\overline{w} \leq 1.$ For our purposes, we make the simplifying assumption the equation of state is functionally constant, or at least that we can approximate it as holding constant at the average value.

To determine the viability of inflationary models, we must relate present day observations to the dynamics of the inflaton field during the inflationary period. This is done by relating a comoving scale, $k$, observed today, to the point during inflation when fluctuations on that scale exited the horizon, defined by 
\begin{equation} 
N_{k} \equiv \ln{\left(\frac{a_{\mathrm{end}}}{a_k}\right)}.
\label{Nkdef}
\end{equation}
To find the relationship between scale $k$ and $N_k$, we begin with the expression relating a given wavenumber $k$ to the size of the sound horizon, $\left(aH\right)$, when it froze out during inflation, $k = \left(aH\right)_k$. We rewrite this as 
\begin{equation}
    \ln{\left(\frac{k}{a_0H_0}\right)}=\ln{\left[\frac{k}{\left(aH\right)_k}\frac{\left(aH\right)_k}{a_0H_0}\right]}=\ln{\left(\frac{\left(aH\right)_k}{a_0H_0}\right)},
\end{equation}
expanding the log to cover the various evolutionary epochs as 
\begin{equation}
     \ln{\left(\frac{k}{a_0H_0}\right)}= \ln{\left(\frac{\left(aH\right)_k}{\left(aH\right)_{\mathrm{end}}}\right)}+\ln{\left(\frac{\left(aH\right)_{\mathrm{end}}}{\left(aH\right)_{\mathrm{RD}}}\right)}+\ln{\left(\frac{\left(aH\right)_{\mathrm{RD}}}{\left(aH\right)_{\mathrm{eq}}}\right)}+\ln{\left(\frac{\left(aH\right)_{\mathrm{eq}}}{\left(aH\right)_{\mathrm{0}}}\right)},
     \label{epochexpand}
\end{equation}
where the subscript ``end'' represents the value at the end of inflation, the subscript ``RD'' is equivalent to ``re'', indicating the value at the end of reheating/the beginning of radiation domination, and ``eq'' indicates the value at matter-radiation equality.  Assuming a constant equation of state, we next use the identity 
\begin{equation}
    aH \propto a^{-(1+3w)/2},
\end{equation} 
and the definition of $N_k$ given in Eq. \ref{Nkdef}.
Substituting these two identities into \ref{epochexpand}, we have 
\begin{equation}
    \ln{\left(\frac{k}{a_0H_0}\right)}=-N_k+ \ln{\left(\frac{H_k}{H_{\mathrm{end}}}\right)}+\left(\frac{1+3\overline{w}}{2}\right)N_{\mathrm{re}}-N_{\mathrm{RD}}+\ln{\left(\frac{\left(aH\right)_{\mathrm{eq}}}{\left(aH\right)_{\mathrm{0}}}\right)}.
    \label{Nk1}
\end{equation}
Here, $N_{\mathrm{RD}}$ is the number of e-folds of expansion between the onset of radiation domination and matter-radiation equality. To evaluate this, we rewrite it as 
\begin{equation}
    N_{\mathrm{RD}}=\ln\left(\frac{a_{\mathrm{RD}}}{a_{\mathrm{eq}}}\right)=-\ln{\left(\frac{T_{re}}{T_{\mathrm{eq}}}\right)}-\frac{1}{3}\ln{\left(\frac{g_{*S}\left[T_{re}\right]}{g_{*S}\left[T_{\mathrm{eq}}\right]}\right)}.
\end{equation}
From the Planck values for the matter and photon densities, \cite{Aghanim:2018eyx}, we can write the redshift of matter/radiation equality as
\begin{equation}
    1+z_{\mathrm{eq}}=\left(\frac{a_0}{a_{\mathrm{eq}}}\right)=\left(\frac{T_{\mathrm{eq}}}{T_{\mathrm{0}}}\right)=\left(\frac{\Omega_mh^2}{\Omega_{\gamma}h^2}\right)=3404,
\end{equation}
so that 
\begin{equation}
    T_{\mathrm{eq}}=3404T_0=9295\mathrm{K} =8.01\times 10^{-10}\ \mathrm{ GeV}.
\end{equation}
We next derive an equation for $\left(aH\right)_z / \left(aH\right)_{\mathrm{0}}$ from the Friedmann equation, $H^2 / H_0^2=\Omega_ma^{-3}+\Omega_{\gamma}a^{-4}+\Omega_{\Lambda},$ 
\begin{equation}
    \frac{\left(aH\right)^2_z}{\left(aH\right)^2_0}=\frac{1}{(1+z)^2}\left[\Omega_{m0}(1+z)^3+\Omega_{\gamma 0}(1+z)^4+\Omega_{\Lambda 0}\right].
\end{equation}
Taking $z=z_{\mathrm{eq}}=3404$, $\Omega_{m0}=0.3166,$ $\Omega_{\gamma 0}=9.32\times 10^{-5},$ and $\Omega_{\Lambda 0}=1-\Omega_{m 0}-\Omega_{\gamma 0}=0.6833,$ we find
\begin{equation}
    \ln\left(\frac{\left(aH\right)_{\mathrm{eq}}}{\left(aH\right)_0}\right)=3.839.
\end{equation}
 Plugging these values into \ref{Nk1}, we find a straightforward equation for $N_k,$ 
\begin{equation}
    N_k=-\ln{\left(\frac{k}{a_0H_0}\right)}+ \ln{\left(\frac{T_{\mathrm{re}}}{10^{25}\  \mathrm{ eV}}\right)}+\frac{1}{3}\ln{\left(\frac{g_{*S}\left[T_{re}\right]}{g_{*S}\left[T_{\mathrm{eq}}\right]}\right)}+\left(\frac{1+3\overline{w}}{2}\right)N_{\mathrm{re}}+\ln{\left(\frac{H_{k}}{H_{\mathrm{end}}}\right)}+61.62,
    \label{Nk2}
\end{equation}
an equation which explicitly shows the dependence of $N_k$ on the reheating epoch. Since the values of $r$, $n_S$, and the normalization of the potential (and therefore  $V_{\mathrm{end}}$) all depend on evaluating certain inflationary parameters at the point when the pivot-scale modes froze out, changing $N_k$ changes that evaluation point, and results in shifting all of these, shifts which need to be taken into account when evaluating the observable parameters of inflationary models. 

\subsection{The effects of $\overline{w}$ and $T_{\mathrm{re}}$. }
In this section, we discuss how $\overline{w}$ influences $N_k$ (and consequently $r$ and $n_S$) in terms of two cases: $\overline{w}>1/3$ and $\overline{w}<1/3$. This distinction reflects the fact that $\overline{w}=1/3$ corresponds to instantaneous reheating followed by radiation domination, as it implies that we have radiation domination for the entire period between the end of inflation and BBN. The cases $\overline{w}<1/3$ and $\overline{w}>1/3$ exhibit drastically different behaviors, which require separate discussions. While it does not have any other direct impact, increasing the length of the transition period (by decreasing $T_{\mathrm{re}}$) increases the effect relative to instantaneous reheating by increasing the duration of the period of expansion with equation of state $\overline{w}$. 

To obtain a bound, we take the limit of the longest possible $\overline{w}$-period, so that $T_{\mathrm{re}}=T_{\mathrm{BBN}}$. 
The first equation we use to parametrize this period is: 
\begin{equation}
\dfrac{\rho_{\mathrm{end}}}{\rho_{\mathrm{BBN}}} =\left(\frac{a_{\mathrm{end}}}{a_{\mathrm{BBN}}}\right)^{-3\left(1+\overline{w}\right)}.
\label{writ1}
\end{equation}
We further take equations for $\rho_{\mathrm{end}}$ and $\rho_{\mathrm{BBN}}$  \cite{Cook:2015vqa}:
\begin{equation}
\rho_{\mathrm{end}}=\frac{3}{2}V_{\mathrm{end}},
\label{writ2}
\end{equation}
and 
\begin{equation}
\rho_{\mathrm{BBN}}=\frac{\pi^2}{30}g_{\mathrm{BBN}}T_{\mathrm{BBN}}^4,
\label{writ2.5}
\end{equation}
where ``$g_{\mathrm{BBN}}$'' represents the number of relativistic degrees of freedom at BBN. 
\begin{figure}
\includegraphics[scale=0.35]{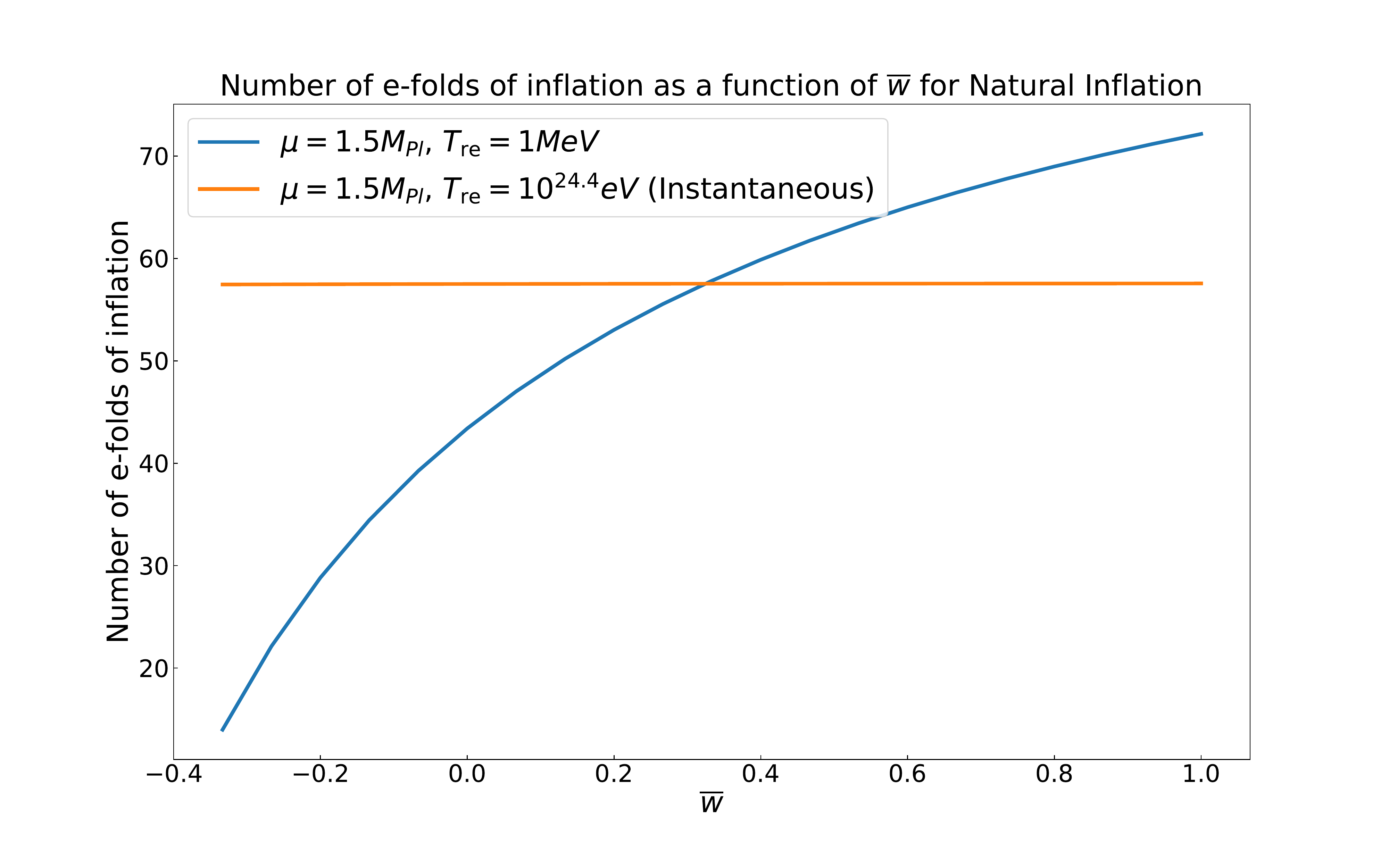}
\caption{Number of e-folds of inflation vs equation of state during $\overline{w}$ period. }
\label{figNNvsw}
\end{figure}
While the value of $V_{\mathrm{end}}$ depends on both the model and the number of e-folds of inflation, for simplicity we first hold it constant, so we can substitute these into Equation (\ref{writ1}). For ease of notation, we define a parameter $\Gamma \equiv \rho_{\mathrm{end}} / \rho_{\mathrm{BBN}}$, and combine Eqs. (\ref{writ1}), (\ref{writ2}), and (\ref{writ2.5}): 
\begin{equation}
\dfrac{\rho_{\mathrm{end}}}{\rho_{\mathrm{BBN}}} =\dfrac{\frac{3}{2}V_{\mathrm{end}},}{\frac{\pi^2}{30}g_{\mathrm{BBN}}T_{\mathrm{BBN}}^4.}=\left(\frac{a_{\mathrm{end}}}{a_{\mathrm{BBN}}}\right)^{-3\left(1+\overline{w}\right)}=\Gamma.
\label{writ3}
\end{equation}
Meanwhile, we know that $\left(aH\right)^{-1}\propto a^{\left(1+3w\right)/2}$, so that we can define the change in the size of the comoving horizon during the transition period as 
\begin{equation}
\frac{\left(aH\right)^{-1}|_{\mathrm{BBN}}}{\left(aH\right)^{-1}|_{\mathrm{end}}}=\left(\frac{a_{\mathrm{BBN}}}{a_{\mathrm{end}}}\right)^{\frac{1}{2}\left(1+3\overline{w}\right)}.
\label{writ4}
\end{equation}
By combining equations \ref{writ3} and \ref{writ4}, we write the change in $\left(aH\right)^{-1}$ during the $\overline{w}$ period as 
\begin{equation}
\frac{\left(aH\right)^{-1}|_{\mathrm{BBN}}}{\left(aH\right)^{-1}|_{\mathrm{end}}}=\Gamma^{\frac{1+3\overline{w}}{6\left(1+\overline{w}\right)}}\equiv\Gamma_{\overline{w}},
\end{equation}
which we can compare to instantaneous reheating by taking the ratio of $\Gamma_{\overline{w}}$ to $\Gamma_{\gamma}$ (the value of $\Gamma$ assuming instantaneous reheating and subsequent radiation domination):
\begin{equation}
\frac{\Gamma_{\overline{w}}}{\Gamma_{\gamma}}=\Gamma^{\frac{1}{6}\left(\frac{1+3\overline{w}}{1+\overline{w}}-\frac{3}{2}\right)}.
\end{equation}
Examining this expression, we can see that, for any fixed value of $\Gamma$, having $\overline{w}>1/3$ leads to \textit{increased} change in the size of the horizon between the end of inflation and Big Bang Nucleosynthesis, whereas $\overline{w}<1/3$ results in a \textit{reduced} change in the size of the horizon during this period. This explains the impact $\overline{w}$ has on $N_k$, and consequently $n_s$ and $r$: our reconstruction of the horizon size at the end of inflation depends on the evolution of the horizon between the end of inflation and BBN, and relate  to match the pivot scale, $\overline{w}>\frac{1}{3}$ requires \textit{more} e-folds, while  $\overline{w}<\frac{1}{3}$ requires \textit{fewer} e-folds, as shown in Figure \ref{figNNvsw}. 

\FloatBarrier
\subsection{Methodology}
In order to exactly calculate the $r$ and $n_S$ values generated by each combination of inflationary potential and set of reheating parameters, we created a Mathematica code which takes an un-normalized inflationary potential and a corresponding mass scale $\mu$ (in this case, $V(\theta)=1+\cos{\theta}$, where $\theta \equiv \phi / \mu$) and the two reheating parameters (average equation of state during reheating, $\overline{w}$, and the temperature at the onset of radiation domination, $T_{\mathrm{re}}$) and outputs predicted values of $N_k$, $r$ and $n_S$.

For the first step, the desired output is the potential at the end of inflation ($V_{\mathrm{end}}$) and $H_k$, both of which depend on the number of e-folds of inflation ($N_k$). The formula for $V_{\mathrm{end}}$ can be written in terms of $\theta_k$ and $\theta_{\mathrm{end}}$, where $\theta_{\mathrm{end}}$ is the value of the field at the end of inflation, and $\theta_k$ is the value at $N_k$ e-folds, as:
\begin{equation}
    V_{\mathrm{end}} = \frac{3\mathrm{A_s}M^2}{128 \pi}\left(\frac{V'\left(\theta_k\right)}{V\left(\theta_k\right)}\right)^2\left(\frac{V\left(\theta_{\mathrm{end}}\right)}{V\left(\theta_k \right)}\right).
    \label{Vend}
\end{equation}
\cite{Freese:2014nla}
Here $M \equiv m_{\mathrm{Pl}} / \mu$, $\mathrm{A_s}$ is the initial amplitude of the pivot-scale curvature fluctuations (taken here to be $2.105\times 10^{-9}$, in accordance with the most recent Planck results, \cite{Aghanim:2018eyx}), and $\theta_{\mathrm{end}}$ is found by numerically solving the equation for $\epsilon(\theta_{\mathrm{end}})=1,$ while  $\theta_k$ is found by numerically solving
\begin{equation}
    N_k = \frac{8 \pi}{M^2}\int_{\theta_k}^{\theta_{\mathrm{end}}}\left[\frac{V(\theta)}{V'(\theta)}\right]\mathrm{d}\theta
\end{equation}
for $\theta_k$ given a certain value of $N_k$. Similarly, $\theta_{\mathrm{end}}$ is found by numerically solving the equation for $\epsilon(\theta_{\mathrm{end}})=1.$ (Here a prime indicates a derivative taken with respect to $\theta,$ rather than $\phi.$)

We write the equation for $H_k$ in terms of the same parameters, as
\begin{equation}
H_k = m_{\mathrm{Pl}}\sqrt{\frac{\mathrm{A_s}}{16}}\left(\frac{V'\left(\theta_k\right)}{V\left(\theta_k\right)}\right).
\label{Hk}
\end{equation}
By rearranging equation 2.4 from \cite{Cook:2015vqa} for $N_{\mathrm{re}}$,
\begin{equation}
    N_{\mathrm{re}}=\frac{1}{3(1+\overline{w})}\ln{\frac{45V_{\mathrm{end}}}{\pi^2g_{\mathrm{*S}}\left(T_{\mathrm{re}}\right) T_{\mathrm{re}}^4}},
    \label{Nre1}
\end{equation} and combining it with equations \ref{Nk2}, \ref{Vend} and \ref{Hk}, we find an equation for $T_{\mathrm{re}}$:
\begin{eqnarray}
    T_{\mathrm{re}} =&&\left(\frac{45V_{\mathrm{end}}}{g_{\gamma}\pi^2}\right)^{1/4} \exp\left[\frac{3\left(1+\overline{w}\right)}{2(1+3\overline{w})}\right] \times \cr 
    &&\exp\left[N_k-\ln{\frac{T_{\mathrm{re}}}{10^{25}\ \mathrm{eV}}}-\frac{1}{3}\ln{\left(\frac{g_{*S}\left(T_{re}\right)}{g_{*S}\left(T_{\mathrm{eq}}\right)}\right)} +\ln{\frac{k}{a_0H_0}}-\ln{\frac{H_k}{H_{\mathrm{end}}}}-61.62\right],
    \label{Tre}
\end{eqnarray}
which we finally solve numerically to find the corresponding value of $N_k$ for the specified initial value parameters, giving us all the information we require to calculate the actual observables, $r$ and $n_s$.

We compare the model predictions to the regions of the $r$ / $n_s$ plane which are allowed by the Planck 2018 TT/TE/EE temperature and polarization data~\cite{Aghanim:2018eyx,Aghanim:2019ame,Akrami:2019bkn,Akrami:2019izv,Akrami:2018odb}, and the BICEP2/Keck Array 2015 combined polarization data~\cite{Ade:2015tva}. We include Baryon Acoustic Oscillation (BAO) data from the Sloan Digital Sky Survey Data Release 12 \cite{Alam:2015mbd}, the 6DF Data Release 3 \cite{Jones:2009yz}, and the Sloan Digital Sky Survey Data Release 7 main galaxy sample (MGS) sample \cite{Ross:2014qpa}. The allowed regions are calculated numerically the \texttt{CosmoMC} Markov Chain Monte Carlo (MCMC) sampler~\cite{Lewis:2002ah}, and the CAMB Boltzmann code. We fit to a seven-parameter $\Lambda$CDM+$r$ model with the following parameters:
\begin{itemize}
\item{Baryon density $\Omega_{\rm b} h^2$.}
\item{Dark matter density $\Omega_{\rm C} h^2$.}
\item{Angular scale of acoustic horizon $\theta$ at decoupling.}
\item{Reionization optical depth $\tau$.}
\item{Power spectrum normalization $A_s$.}
\item{Tensor-to-scalar ratio $r$, calculated at a pivot scale of $k = 0.05\ h \mathrm{Mpc}^{-1}$.}
\item{Scalar spectral index $n_{\rm S}$.}
\end{itemize}
In our analysis, we assume curvature $\Omega_{\rm k}$ is zero, and the Dark Energy equation of state is $w = -1$. We set the number of neutrino species $N_\nu = 3.046$, and neutrino mass $m_\nu = 0.06\ {\rm eV}$. We apply Metropolis-Hastings sampling to 8 chains running in parallel, and use a convergence criterion for the Gelman-Rubin parameter $R$ of $R - 1 < 0.05$. 

\FloatBarrier
\section{Results}
\subsection{Case 1: Conventional Post-Inflationary Dynamics}
\begin{figure}
\centering
\includegraphics[width=0.9 \textwidth]{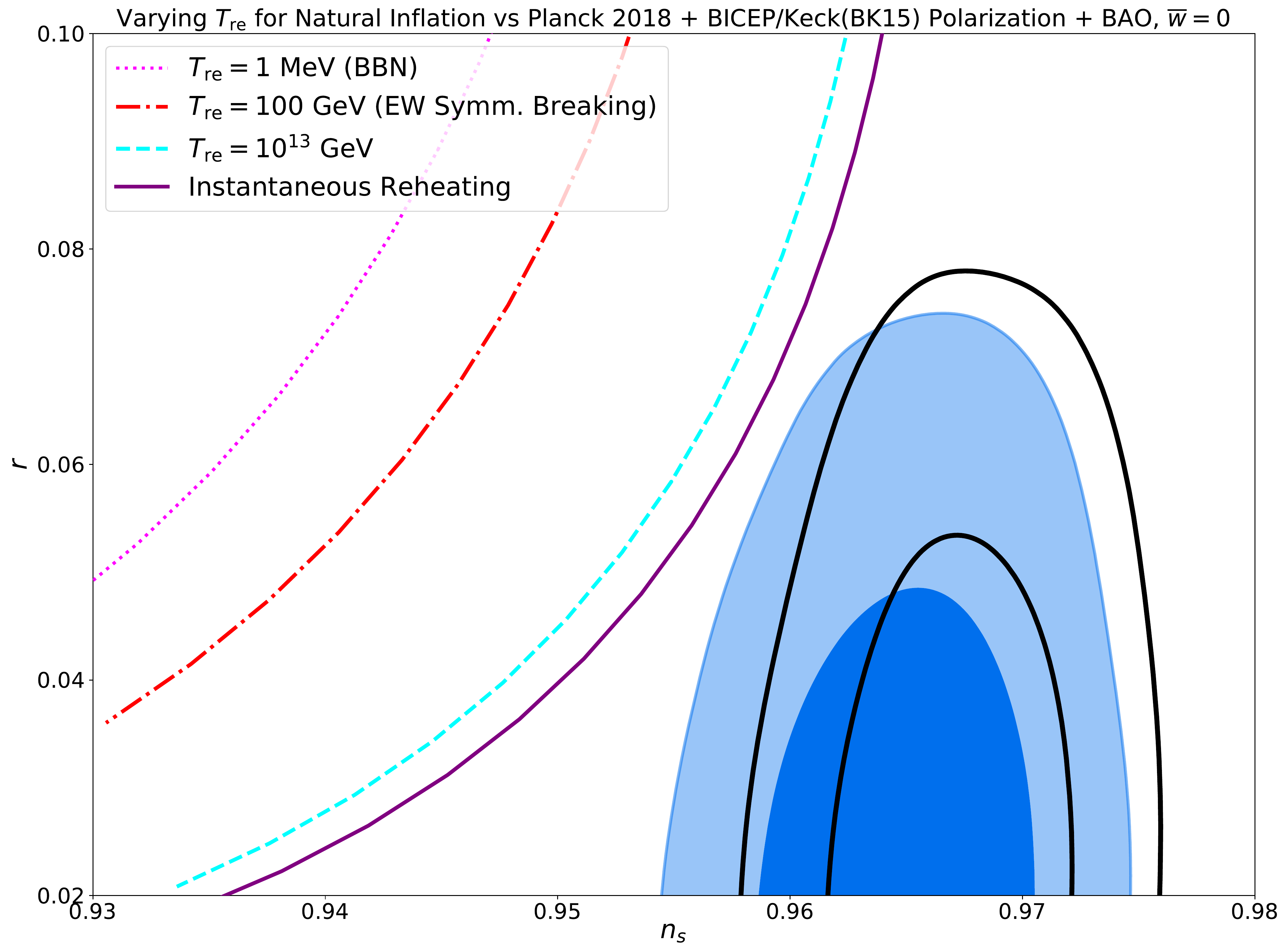}
\caption{ Tensor/scalar ratio $r$ vs spectral index $n_s$ for a variety of values of $T_{\mathrm{re}}$, assuming  $\overline{w}=0$. The blue shaded region represents the allowed region for Planck 2018 + BICEP/Keck(BK15) Polarization + BAO, while the black line contours represent the addition of the SHOES $H_0$ data. (Given the statistically significant tension between the SHOES constraint on $H_0$ and the constraint from Planck, combining the two in a Bayesian fit is likely of limited value. We include the SHOES constraint only to show that our conclusion is not sensitive to the assumed value of $H_0$.) Instantaneous reheating comes closest to matching observations, but is still excluded from the 95\% confidence region. Decreasing $T_{\mathrm{re}}$ only increases the tension, moving the $r-n_s$ curves up and to the left, away from the allowed region. }
\label{FigTvarref}
\end{figure}
We first consider the simple case of conventional reheating directly to a radiation-dominated universe after inflation.
Since, near the minimum, the mass term dominates the potential, the average equation of state during reheating is $\overline{w}=0$, though the duration of reheating can vary. This case, shown in Fig. \ref{FigTvarref}, is disfavored at greater than 95\% confidence, and decreasing $T_{\mathrm{re}}$ decreases the number of e-folds of inflation, exacerbating the inconsistency with data. Note that in the limit of instantaneous reheating, $\overline{w}$ becomes irrelevant; the instantaneous reheating line on figures \ref{FigTvarref} and \ref{FigTvar} are identical, and equivalent to the scenario where $\overline{w}=1/3$.

\subsection{Case 2: General Post-Inflationary Dynamics}
\begin{figure}
\centering
\includegraphics[width=0.9 \textwidth]{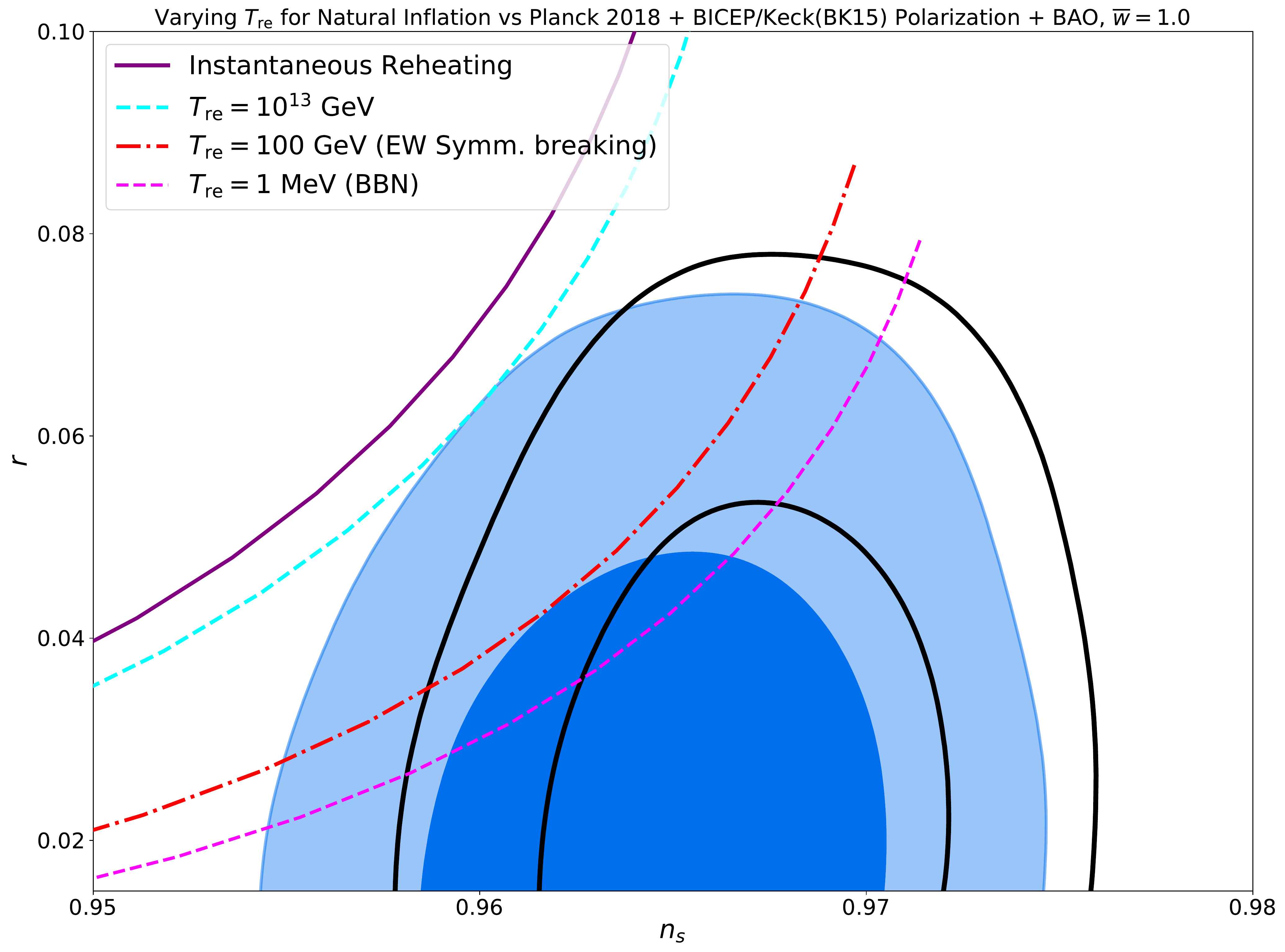}
\caption{Curves in the $r-n_s$ plane for a variety of values of $T_{\mathrm{re}}$ in the limit of $\overline{w}=1$. The blue shaded region represents Planck 2018 + BICEP/Keck(BK15) Polarization + BAO, while the black line contours represent the addition of the SHOES $H_0$ data. Note that, although the range in $n_s$ is different, the curve representing instantaneous reheating here is exactly equivalent to its counterpart on Figure \ref{FigTvarref}.}
\label{FigTvar}
\end{figure}

\begin{figure}
\centering
\includegraphics[width=0.9 \textwidth]{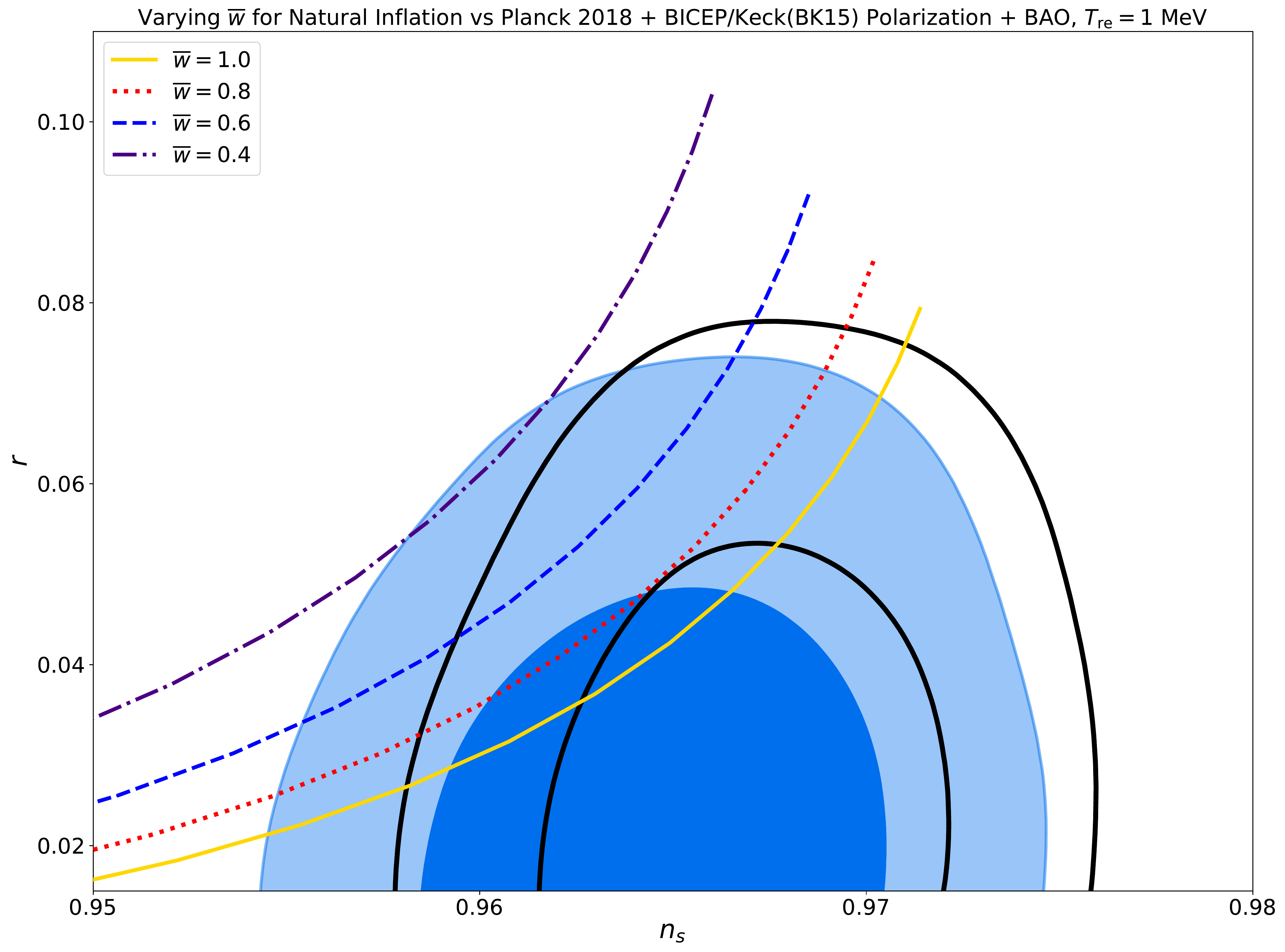}
\caption{Curves in the $r-n_s$ plane for a variety of values of $\overline{w}$  in the limit of $T_{\mathrm{re}}=1\ \mathrm{MeV}$. The blue shaded region represents Planck 2018 + BICEP/Keck(BK15) Polarization + BAO, while the black line contours represent the addition of the SHOES $H_0$ data. }
\label{Figwvar}
\end{figure}

While a cosine potential results in an equation of state during reheating of $\overline{w}=0$, `reheating' doesn't necessarily have to be the only thing that occurs during the $\overline{w}$ period. If we posit some other interaction or field domination between the end of inflation and BBN, either during or after the decay of the inflaton field itself, then it could be possible to create a situation where $\overline{w}$ is greater than $1/3$ for an arbitrarily long period between inflation and BBN. If we allow $\overline{w}>\frac{1}{3}$, we can increase $N_{k}$, which shifts the $r-n_s$ curves down and to the right, reducing the tension with observations. Ignoring SHOES, if we take the limit $\overline{w}=1$, we find that Natural Inflation agrees with the 95\% range of current observations for $T_{\mathrm{re}}<10^{13}\ \mathrm{GeV}$, and the 68\% region if we extend the $\overline{w}$ period past the electroweak scale at $100\ \mathrm{GeV}$, as shown in Figure \ref{FigTvar}. Similarly, if we take the limit of $T_{\mathrm{re}}=1\ \mathrm{MeV}$, using a value of $\overline{w}\geq \approx0.38$ brings the $r-n_s$ curve into the 95\% range, and $\overline{w}\geq\approx0.75$ takes us into the 68\% region, as shown in Figure \ref{Figwvar}. However, as shown in Ref. \cite{Tanin:2020qjw}, we must be careful when extending reheating into late times, especially when it represents a `stiff' epoch (i.e. $\overline{w}>1/3$), since the energy density of the gravitational waves is amplified during such a period, and we can tightly constrain the stochastic GW background at BBN.

\FloatBarrier
\section{Conclusions}
In this paper, we have shown that the original model of Natural Inflation using a cosine potential is inconsistent with constraints on $r$ and $n_s$ at greater than 95\% confidence. However, if we allow for unconventional reheating characterized by a period of $\overline{w}>1/3$, Natural Inflation can be brought into agreement with current observations. While it is difficult to bring these models into within the 68\% confidence region without extending reheating into temperatures between electroweak symmetry breaking and nucleosynthesis, it is entirely possible to bring them into the 95\% confidence region, and by extending reheating below the electroweak scale, Natural Inflation with nonstandard reheating can generate $r-n_s$ values well within the 68\% confidence range, down to a minimum $r$ value of $r \sim \mathcal{O}(0.01)$. However, if future CMB experiments fail to detect tensor modes on that scale, even this extension of the parameter space accessible to Natural Inflation models will be insufficient, and this choice of scalar field potential will be ruled out entirely. 

\section*{Acknowledgments}
This work is supported by the National Science Foundation under grants NSF-PHY-1719690 and NSF-PHY-2014021.  This work was performed in part at the University at Buffalo Center for Computational Research.

\FloatBarrier
\bibliographystyle{JHEP}
\bibliography{NatInflPaper.bib}
\end{document}